# Developing a Novel Crowdsourcing Business Model for Micro-Mobility Ride-Sharing Systems: Methodology and Preliminary Results

**Mohammed Elhenawy**
Research fellow at Centre for Accident Research and Road Safety
Queensland University of Technology, Australia
mohammed.elhenawy@qut.edu.au

**MD Mostafizur Rahman Komol**
Researcher at Centre for Accident Research and Road Safety
Queensland University of Technology, Australia
mdmostafizurrahman.komol@hdr.qut.edu.au

**Huthaifa I. Ashqar (Corresponding Author)**
Booz Allen Hamilton
Washington, D.C., 20003 USA
hiashqar@vt.edu

**Mohammed Hamad Almannaa**
Assistant professor, Civil Engineering Department
King Saud University, Riyadh, Saudi Arabia
malmannaa@ksu.edu.sa

**Mahmoud Masoud**
Research associate at the School of Mathematical Sciences
Queensland University of Technology, Australia
mahmoud.masoud@qut.edu.au

**Hesham A. Rakha**
Samuel Reynolds Pritchard Professor of Engineering and Director of the Center for Sustainable Mobility
Virginia Tech Transportation Institute, Blacksburg, Virginia 24060
hrakha@vt.edu

**Andry Rakotonirainy**
Director of the Centre for Accident Research and Road Safety
Queensland University of Technology, Australia
r.andry@qut.edu.au

Word Count: 6,335 words + 4 table (250 words per table) = 7,335 words

## Abstract

Micro-mobility ride-sharing is an emerging technology that provides access to the transit system with minimum environmental impacts. Significant research is required to ensure that micro-mobility ride-sharing provides a better fulfilment of user needs. In this study, we propose a novel business model for the micro-mobility ride-sharing system where light vehicles such as electric scooters and electric bikes are crowdsourced. This new model consists of three entities, the suppliers, the customers, and a management party, which is responsible for receiving, renting, booking, and demand matching with offered resources. The proposed model has the potential to allow the suppliers to define the location of their private e-scooter/e-bike and the period of time they are available for rent, match it with a particular demand, and then offer suppliers the opportunity to get their e-scooters/e-bikes rented and returned at the end of the renting period to the same (nearby) location. The management party will need to match the e-scooter/e-bike to a series of renting demands with the last demand as a destination very close to the initial location of the e-scooter/e-bike at the start of the renting period. One potential advantage of the proposed model is that it shifts the charging and maintenance efforts to a crowd of suppliers.

## Introduction

At present days, micro-mobility is a promising urban mobility solution [1]. The term micro-mobility deals with the incorporation of a short-trip by small vehicle operation. When transportation mobility is restrained to a very limited range of trip for light vehicles only, it is called micro-mobility [2]. Vehicles of light categories such as bicycles, motorbike, electric bikes (e-bikes), electric scooters (e-scooters), shared bicycles, and some riding devices like skateboards are considered as micro-mobility rides [3]. Micro-mobility ride sharing is a marvelling technique in providing sustainability at this current hazards of over-growing population, traffic congestion and greenhouse impact [4]. Environment pollution and Carbon dioxide emission can be minimised to an enormous amount of utilising micro-mobility rather than regular motor vehicles [5, 6]. Recently, the public interest on micro-mobility like bike-sharing systems (BSS) or scooters-sharing systems (SSS) is proliferating with the rapid transformation of the transportation system. Especially for traffic congested routes, this strategy gives an affordable and rapid passage opportunity, saving civilians from intolerable waiting and the squander of time [7-9].

Micro-mobility ride-sharing system is categorised as four generations [10] where dock-based systems are considered as latest updated versions of the first two generations [11, 12]. The emergence of fourth-generation dockless ride-sharing has brought a revolution in the micro-mobility ride-sharing market [13] due to mass public deliberation to the sharing economy, sustainability and health transports [14]. The early invention of micro-mobility ride-sharing strategy was initiated as dockless bike-sharing, but this system was not prioritised enough due to the lack of technologies to fulfil the system drawbacks. People rather accepted dock-based ride-sharing as more flexible and reliable in comparison to the dockless ride-sharing. However, in modern days, the new available technologies have assured the dockless ride-sharing far more flexible and reliable to users expectation [15]. Moreover, this system is handy on installation, less intricate for smartcard integration system on transit and facilitated with advanced power assisting features [10]. Rather the dock-based system causes an extra infrastructure cost to be paid for the agglomeration and locking security of micro-mobility rides [15]. Nonetheless, the dockless system is considered as more flexible than the dock-based system, their advantages and superiority of implementation varies based on the feature of locations [16] and also on the purpose of the trip, how fast user walks, destination, budgets for the rides, reliability value and any other operators attributes [17]. Even the dockless micro-mobility system is bound to a lot of limitations. Many of such businesses got collapsed like a short flash of fireworks due to the hardship in maintaining sustainability, oversized fleets and vandalism [15, 18, 19]. Moreover, in dockless system, range of the distance to be travelled by scooters is needed to be restricted due to the limitation of battery power charge of scooters. Otherwise, the commuters will requisite to abandon the shared scooter in the halfway to their destination if the charge drops below usability condition [16]. In some European countries, private e-scooter is overgrowing rapidly [20] and shared scooter policy is being observed in withdrawal eyes in those places. For extensive use of micro-mobility rides in public roads, it often causes annoyance and conflict with the safety of regular pedestrians [21]. That is why there is also a profusion of complaining over the exaggeration of light vehicles which impact negatively over the business idea of micro-mobility ride sharing.

So, the clarification of ambiguity of previous inventions as well as evaluating novel ideas on the micro-mobility ride-sharing system is a pivotal research focus of the modern era. Considering all the challenges and hurdling experiences with conventional micro-mobility ride sharing systems, a massive improvement and modifications are desired to cover the drawbacks pertaining to this field. Still, a lot of further research is required to give the micro-mobility ride-sharing system a better fulfilment of users' expectation. The innovation of business models like micro-mobility ride-sharing is also getting popular concern and being welcomed in public appearance if it can fulfil challenges confronted by the dock-based and dockless ride-sharing systems. In this research paper, a novel business model is demonstrated over the micro-mobility ride-sharing system focusing on recovering flaws of conventional strategies. The proposed model offers numerous advantages features for all classes included in this business strategy.

Another application to this model is during the periods of pandemics and after the lockdown is ended and restrictions are eased gradually. This model has the potential to serve as a solution to the issues arise after people can return to normal life routines for economic recovery. However, precautions are required to minimize the chance of exposing areas to a second wave of the pandemic. Social distancing is one of the most important counter-measures against the spread of a pandemic, which needs to be in effect for months after the peak. Social distancing is a challenging issue in cities where public transportation is the main commuting means. Consequently, this model could be deployed in some areas of the city to meet some of the demand due to abandoning public transportation.

**Literature Review**

Sharing the micro-mobility rides for public necessity and the business expansion over this strategy is not an unprecedented scheme. This idea first evolved around 1965, but for the lack of technologies to track down customers, the users used to remain anonymous and miserable incidents like stealing and destroying of bikes frequently occurred which led the strategy to collapse for practical implementation [22-24]. As the technology arises with smartphone and GPS system, customers can easily track and locate the nearest available sharable rides through GPS, and they can even unlock them using their smartphones [16, 25]. It is spreading in greater extend from the last decades with the advancement of technology and management policy [26]. So, researchers are driven to put immense effort to improve micro-mobility ride-sharing management and technologies as it is a popular concern and being used ubiquitously.

As time flies, the micro-mobility ride-sharing system is developed and modified with both technology and policy improvements. An innovative simulation study of models without redistribution and with simple redistribution are proposed for more effectual bike redistribution system [27]. When customers intend to rent a bike, they need to know the rental information exploitation mechanisms about the pick-up and the return of bikes. This helps in fulfil their demand properly and prevent probable queue making during renting. In another research, machine learning-based univariate and multivariate regression algorithms have been implemented to model available bikes in rental stations [28-30]. A greedy randomised adaptive search procedure (GRASP) algorithm has also been utilised and improved for bicycle rebalancing problem to solve unavailability of bicycles to rent [31]. Random Forrest (RF) and Multi-Layer perception (MLP) have been used to predict the bike availability in different rental stations using the real-world data of some Swiss cities. Here, the pick-up and the return of the bike is optimised to maintain balance for bike availability based on the prediction [32]. Nowadays, mobile phone apps are also being initiated and developed to locate nearby station to rent and ride availability. Also, route planning and possible parking for commuters are being presented through these apps [33]. For solving ride unavailability in rental stations, attempts are also taken using policymaking. The characteristics of customers in renting rides from different stations have been evaluated. The tendency of choosing an alternative station with less available bikes is found among customers than the preferred ones [34]. Dockless ride-sharing is getting superior with economic advantages in the bike-sharing market by implementing remote mobile payment system and the emergence of big-data computation [8, 15, 35-37]. But for the dockless ride-sharing system, the redistribution due to the imbalance of pick up and the return of rides is more challenging. It is found that maintaining a win-win situation to customers by providing monetary incentives in renting rides is highly effective for balancing the redistribution in the dockless ride-sharing system [38].

E-scooter share trip trajectories were provided at the street link level with precise construction of trajectory trip inventories [39]. Contribution of characters individuals to the reduction of heat alert through the bike-sharing system was analysed in research where the factor of age and gender was also showed different behaviour. Also, different weather and environmental condition and the ability to expand the system [40] are found to be responsible for the variance of micro-mobility ride hiring by users [41, 42]. Apart from age and gender, trip purpose, time of the day, day of the week, population density, median household income and some other demographic and external factors were found as a reagent in other researches on Washington DC and Austin, TX for the variability of ride-sharing [43, 44]. To understand people's perception and opinion over micro-mobility ride-sharing, researchers accumulated data from social media like Twitter, Facebook and utilised machine learning approach to evaluate the current toleration and support of ride-sharing business at the present time [45]. Complexity and Sophistication of mobile applications that operate in micro-mobility ride-sharing assistance are found as an influential factor behind the use of ride-sharing. The young generation is highly acquaintance with complex apps, and they have easily been driven to use micro-mobility as they properly understand the feature and facilities of the system through the apps. However, people of old age highly decline these apps, and they are really sophisticated for them to understand the facility and benefit of this system. For them, an easier version of apps sounds recommended [46].

Considering previous literature as mentioned above, it is found that micro-mobility ride-sharing system is being developed gradually and this system needs a lot of future improvement and modification to solve myriads of challenges that people face every day in operating this business practice. In this research, a new business model for the micro-mobility ride-sharing system is proposed where light vehicles such as electric scooters, electric bikes are crowdsourced. An advantage of this new model is that it shifts the charging and maintenance efforts to the crowd of the suppliers. In conventional ride-sharing system, The maintenance cost and charging cost due to high repositioning rates of shared-scooters is much higher [16], which often gets challenging to afford by customers or business party. But in our proposed strategy, this cost will be covered by suppliers, and both customers and maintenance party will be relieved from the burden of sharing this cost. However, the incentive payments by the customers to suppliers will cover some of its expense as shared policy.

**Proposed Management Policy**

The proposed model consists of three entities. The suppliers, who are the people who are offering their private e-scooters/e-bikes[1] for hire. The customers, who are the people that create hiring demand. Finally, the management party, which is responsible for receiving hiring booking demand and matching it with offered resources. The proposed model allows the suppliers to define the location of their private e-scooter/e-bike and the period of time they are available for rent. The model will allow suppliers to get their e-scooters/e-bikes hired and returned at the end of the renting period to the same location or another near location. This means that the management party needs to match the e-scooter/e-bike to a series of hiring demands with the last demand as a destination very close to the initial location of the e-scooter/e-bike at the start of the hiring period.

An advantage of this new model is that it shifts the charging and maintenance efforts to the crowd of the suppliers. In conventional ride-sharing system, The maintenance cost and charging cost due to high repositioning rates of shared-scooters is much higher [16], which often gets challenging to afford by customers or business party. But in our proposed strategy, this cost will be covered by suppliers, and both customers and maintenance party will be relieved from the burden of sharing this cost. However, the incentive payments by the customers to suppliers will cover some of its expense as shared policy.

There are many approaches that could be used to match the supply and demand in a ride-sharing system. Depending on the booking scheme allowed by the model, we can choose the matching approach. The first possible approach is a data-driven one, which is suitable for real time booking. The naivest variant of the data-driven approach is what can be referred to the two-leg round-trip approach. The current e-scooter service is assumed to be meeting only 50% of the total demand volume, thus; we used the available data about the current trips to estimate the origin-destination matrix of non-observed/hidden demand. Consequently, we plan routes such that we assign e-scooters that are located at route endpoints to service this route. The e-scooters will be assigned trips only over this route such that it travels back and forth between the two ends of the route and the destination of the last trip is the e-scooters location at the starting of the day. This will require the management application to ask the renters to enter the origin and destination of the trip to find them the e-scooter that assigned to this route. Moreover, the management should not allow the e-scooters to get out of their assigned routes. In this approach, the planned routes may vary based on the day-of-the-weak and hour-of-the-day to meet the expected demand. This route variations makes the model flexible and capable of meeting the demand in a more efficient way. Allowing more than two-leg round-trip is more complicated than a data-driven approach. We use the estimated O-D matrix to establish routes between more than two points such that an e-scooter can move freely between this subset of points during its renting period. However, at the end of the rental period, the final destination of the e-scooter should be as closer as possible to its initial location. In this paper, we will use an agent-based modelling (ABM) approach to establish simple management rules which are suitable for the two-leg round-trip approach.

---
[1] Throughout this paper, we occasionally used only *e-scooter(s)* to refer to all other micro-mobility vehicles, as e-scooters are the main vehicles used in the tested dataset in this study.

**Proposed Agent-Based Modelling (ABM)**

In this subsection we discuss the ABM framework of the proposed new model. We established a set of simple rules that will yield a suitable behaviour of the proposed e-scooter model. Our model has three types of agents namely;

1. E-scooter agent,
2. Central agent, and
3. Demand generator agent.

The e-scooter agent has four state variables namely, the home location, the current location, availability, and current battery range. The other two agents do not have state variables. However, the demand generator has one behaviour rule. Every clock tick, the demand generator looks at the trips data and informs all e-scooter agents of new demand requests including trip information. In other terms, if there are concurrent demands, the demand generator agent chooses one of them to broadcast and wait until the demand is processed then broadcasts another one to the concurrent demand requests. The e-scooter agent rule checks if its current location is the same as the origin of the trip, its battery level, and the availability of e-scooters. If all of the above conditions are satisfied, then the e-scooter agent sends an expression of interest (EOI) to the central agent. The central agent receives EIOs to serve a particular demand from all e-scooter agents and chooses the winner based on the system's own rules. In this experimental work, we proposed five rules to choose the winner as follows;

The first rule chooses an e-scooter agent randomly from a subset of e-scooter agents, which is submitted to the EOI.

The second rule chooses the e-scooter agent that submitted the EOI, and it has the largest remaining charge in its battery.

The third rule chooses an e-scooter agent randomly from a subset of e-scooter agents, which submitted EOI, and the destination of the demand (i.e. trip) is the home location of the e-scooter agent.

The fourth rule chooses the e-scooter agent, which submitted EOI, the destination of the demand (i.e. trip) is the home location of the e-scooter agent, and it has the largest remaining charge in its battery.

The fifth rule drops the demand if there is no any EOI.

## Data Analysis and Statistical Methodology

### Dataset

This study used a dataset that was collected in the city of Austin, Texas and is publicly available by the City of Austin [47]. The dataset contains about 9.2 million trips taken by users for either e-bikes or e-scooters from December 2018 to January 2020. Each trip is represented in a row and each of which has 18 features: trip ID, device ID, vehicle type (e-scooter or e-bike), trip duration (in seconds), trip distance (in meter), start time of the trip, end time of the trip, month, hour, day of week, year, council district (both start and end), and census tract (both start and end). Table 1 shows the total number of e-scooter and e-bikes that operate in the city.

Table 1 List of operators licensed to serve in Austin, TX.

| Operator | E-scooters | E-bikes |
|---|---|---|
| Bird | 4500 | 0 |
| JUMP | 2500 | 2000 |
| Lime | 5000 | 0 |
| Lyft | 2000 | 0 |
| OjO | 100 | 0 |
| Skip | 500 | 0 |
| Spin | 500 | 0 |
| VeoRide | 300 | 50 |
| **Total** | **15400** | **2050** |

We should highlight that the data is reidentified by replacing the location of origin and destination of each trip by the census tracts of these locations. We removed all the trips that are missing the start census tract and/or end census tract and consequently constructed the origin-destination (O-D) matrix. The number of trips after removing the above trips reached to 9,174,541 trips (originally it was 9,231,107 trips). Based on the O-D matrix we have two observations. First, most of the trips are concentrated between particular census tracts and secondly, a significant percentage of trips started and ended in the same census tract as shown in Figure 1. Both of those observations are expected because the majority of the trips are e-scooter trips and these trips are usually short trips and e-scooters are deployed in a limited area at the Austin downtown area. To this end, in the following analysis we used a reduced O-D by selecting fourteen census tracts shown in Figure 1 (c) and (d), which include almost 90% of the total trips in Austin, TX. It is apparent that the large numbers on the diagonal of the matrix in Figure 1 (d) means that a high percentage of the trips start and end in the same census tract.

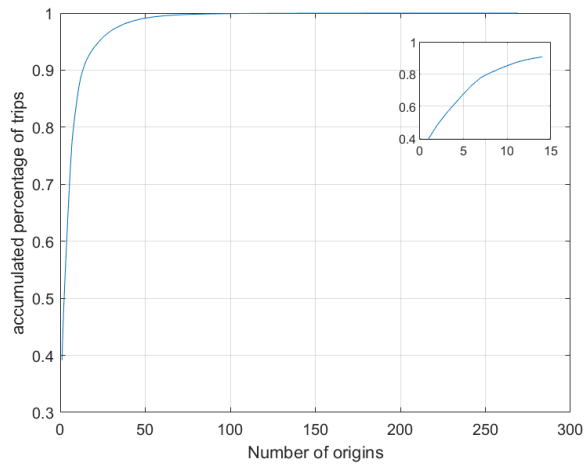

(a)

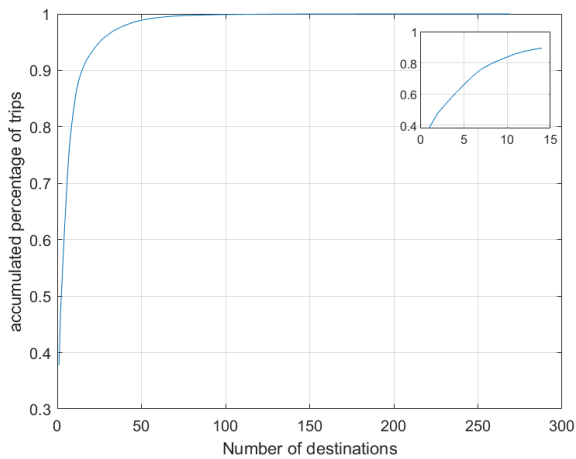

(b)

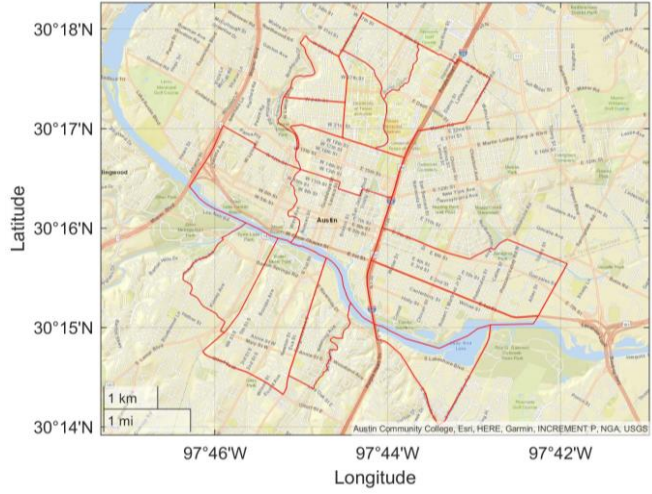

(c)

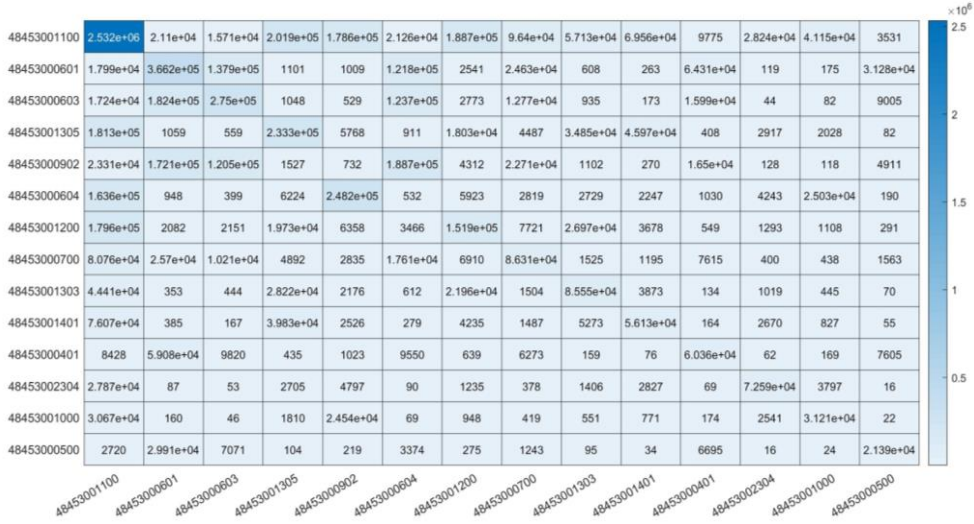

(d)

Figure 1 Most of the trips are concentrated in a contiguous census tracts (a) the accumulated percentage of trips versus number of origins, (b) the accumulated percentage of trips versus number of destinations, (c) the fourteen census tract that include about 90% of the total trips, and (d) the heat map of the O-D matrix of the fourteen census tracts.

**Experimental Results**

In this section, we used the trips data of the reduced O-D matrix to investigate the concept of the new model. In other terms, we used a real dataset of the reduced O-D matrix to run our proposed ABM for different combinations of the central agent rules and chose the combination that satisfies our target performance. We used two criteria to evaluate the aggregated behaviour of the ABM when changing the central agent rules. The first criterion is called Home Index (HI). We defined HI as the percentage of e-scooter agents that ended up at the users' home location at the end of the operation hours, as follows;

$$HI\ (\%) = \frac{\sum_{i=1}^{N} i}{N}$$

where $i$ is an e-scooter that ended up at a user's home at the end of hours of operation, and $N$ is number of the e-scooters that satisfied one demand at least during the hours of operation. The second criterion is the percentage of satisfied demand (PSD), which is the ratio of the demand/trip met at least once during the hours of operation over the total number of trips.

$$PSD\ (\%) = \frac{number\ of\ satisfied\ trips}{total\ number\ of\ trips}$$

We ran the proposed ABM simulation using the 389-day worth of data. We assumed that all e-scooters have the same maximum battery range in the same run. Moreover, we assumed that the operation hours started with the same number of e-scooters at each of the fourteen origins for each run. We simulated the proposed ABM six times using a maximum battery range of (35, 45, and 60 km) and different number of e-scooters at each origin of (50, and 100). HI and PSD were calculated for every day at the different combination of the proposed centre agent rules. To compare the result, we used mixed effect Gamma regression model to explain the variability in HI and PSD in terms of the combination of the proposed rules. Table 2 shows the indicators variables used to encode the tested combination.

Table 2 The indicator variables used to encode the tested combination.

| Scenarios | Rules | $X_1$ | $X_2$ | $X_3$ |
|---|---|---|---|---|
| Scenario 1 | Rule 1 and rule 5 | 0 | 0 | 0 |
| Scenario 2 | Rule 2 and rule 5 | 1 | 0 | 0 |
| Scenario 3 | Rule 3 and rule 5 | 0 | 1 | 0 |
| Scenario 4 | Rule 4 and rule 5 | 0 | 0 | 1 |

The e-scooter trips of the 389 days were used in the ABM environment to simulate the proposed model and estimate the HI for the 389 days. Figure 2 shows the box plots of the HI results versus the different scenarios. The figures show that scenarios 3 and 4 yield higher HI than the other two scenarios, as it includes the rule of which destination of the demand (i.e. trip) is the home location of the e-scooter agent.

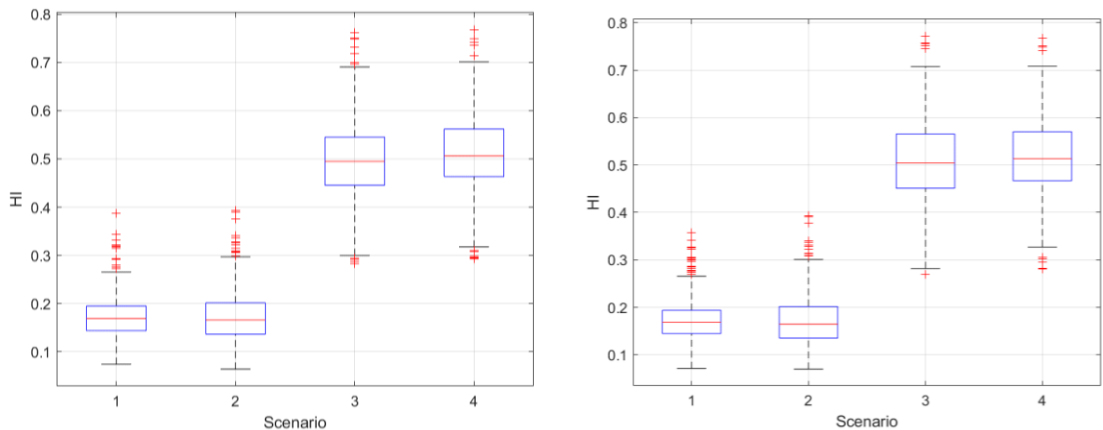

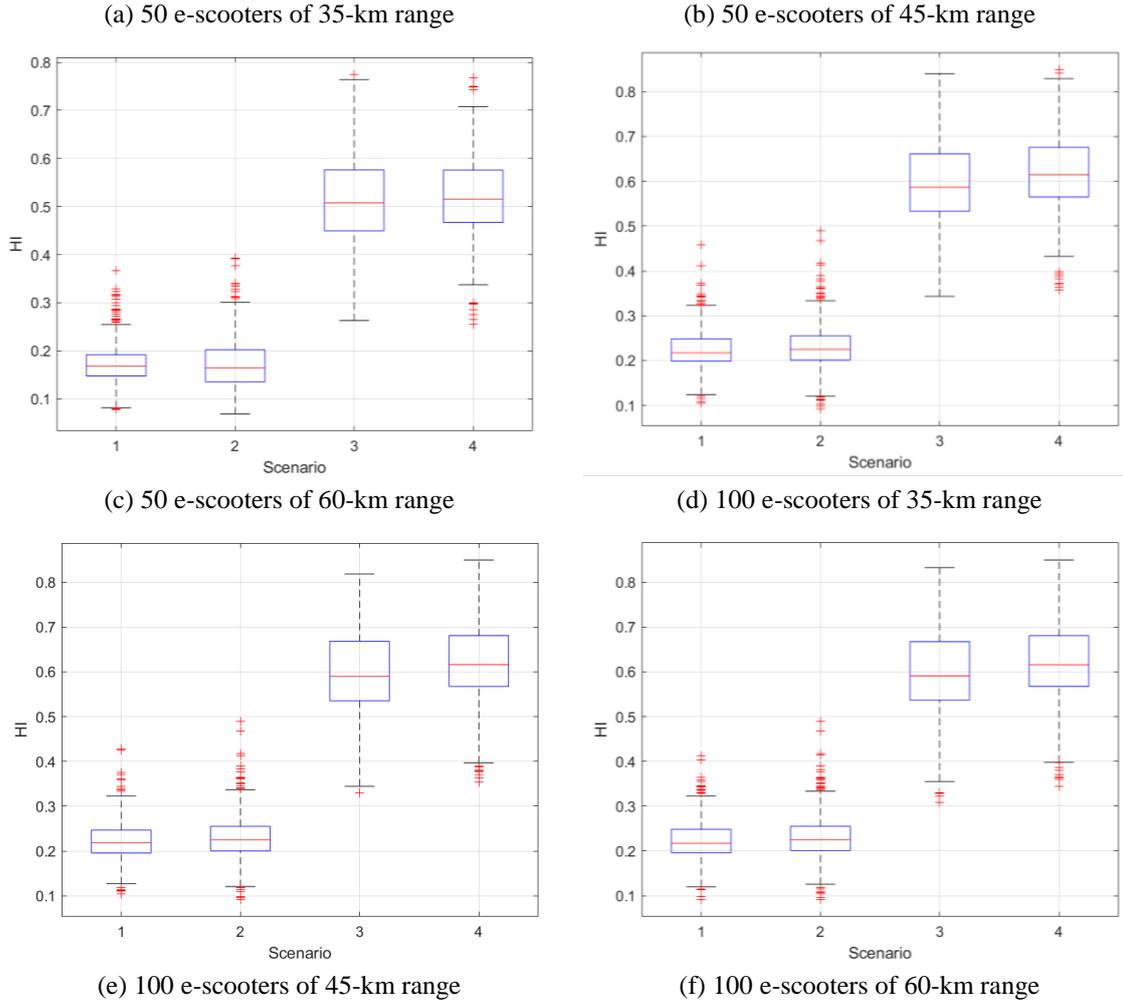

Figure 2 HI results at the different scenarios, number of e-scooters at each census tract, and battery ranges.

Moreover, we used the Gamma mixed effect model to test whether the differences in HI results between the scenarios are statistically significant. As shown in Table 3, the indicator variables corresponding to scenarios 3 ($X_2$) and 4 ($X_3$) are significant. However, only when the system contains 100 e-scooters, the indicator variable for scenario 2 is significant. Results of HI show that when the system contains 50 e-scooters, HI of scenarios 3 and 4 are significantly larger than the HI of the scenario 1. It also shows that at the same condition of 50 e-scooters, HI of scenario 2 is not statistically significant. Finally, it shows that at 100 e-scooters condition, HI of scenarios 2, 3, and 4 are significantly higher than HI of scenario 1.

Table 3 The p-values of the scenario's indicator variables.

| E-scooters number and battery ranges | $X_1$ | $X_2$ | $X_3$ |
| --- | --- | --- | --- |
| 50 e-scooters of 35-km range | 0.67111 | 0 | 0 |
| 50 e-scooters of 45-km range | 0.70196 | 0 | 0 |
| 50 e-scooters of 60-km range | 0.64689 | 0 | 0 |
| 100 e-scooters of 35-km range | 0.00899 | 0 | 0 |
| 100 e-scooters of 45-km range | 0.00255 | 0 | 0 |
| 100 e-scooters of 60-km range | 0.00304 | 0 | 0 |

We next compare between the four scenarios in terms of the percentage of satisfied demand (PSD) at different conditions of the e-scooters number and battery ranges. Figure 3 shows the box plots of the PDS versus the scenarios. The figures show that almost all scenarios have the same PSD values at the same conditions and PSD increases as the used number of e-scooters in the model increases at each census tract. We also modelled PSD results using mixed effect Gamma regression model as a function in the scenarios. The p-values of the indicator variables corresponding to the scenarios are shown in Table 4. These p-values dictate that scenario 3 is different from the scenario 1 at all conditions.

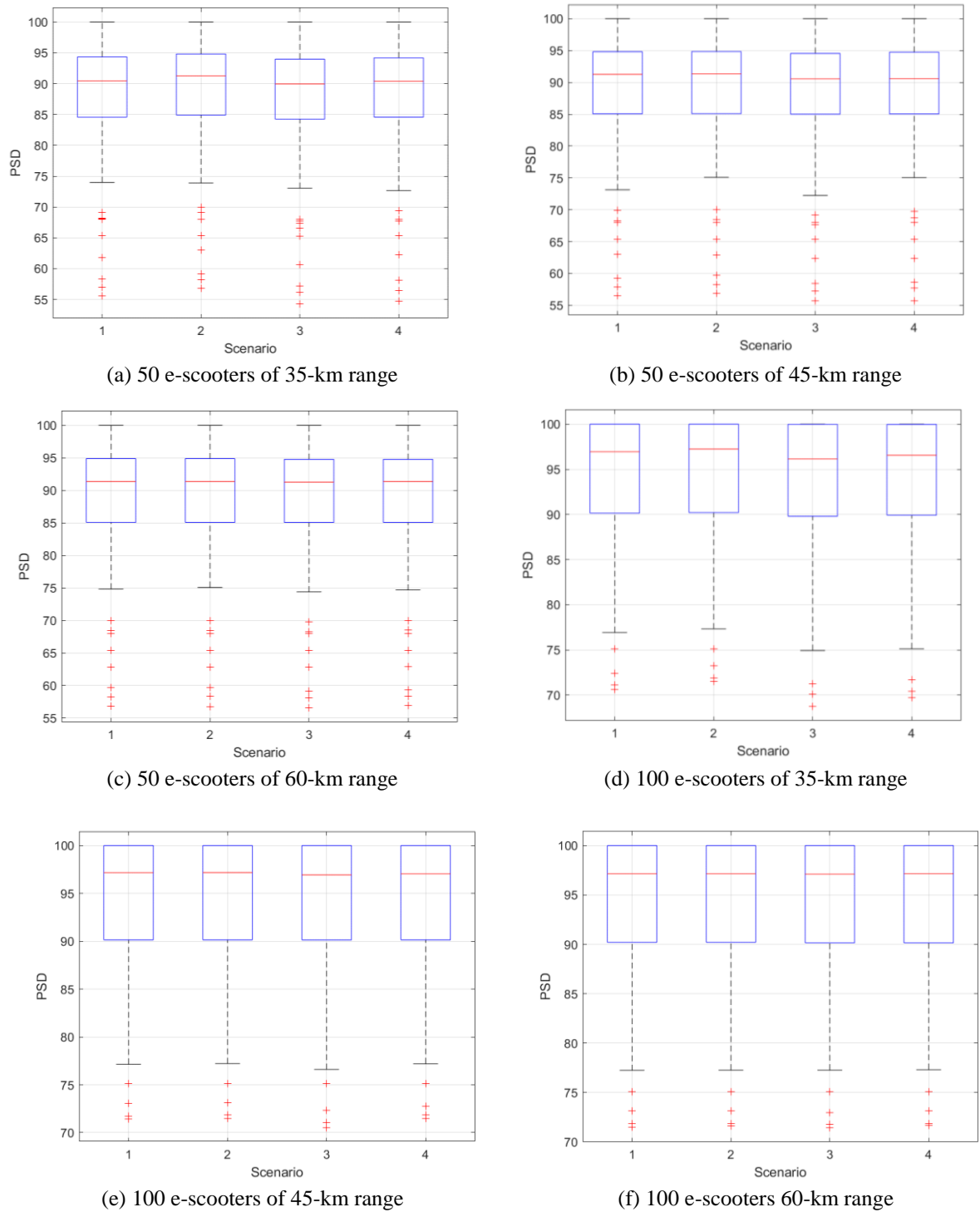

(a) 50 e-scooters of 35-km range  (b) 50 e-scooters of 45-km range

(c) 50 e-scooters of 60-km range  (d) 100 e-scooters of 35-km range

(e) 100 e-scooters of 45-km range  (f) 100 e-scooters 60-km range

Figure 3 PSD results at different scenarios, number of e-scooters at each census tract, and battery ranges.

Table 4 The p-values of the scenario's indicator variables

| E-scooters number and battery ranges | $X_1$ | $X_2$ | $X_3$ |
|---|---|---|---|
| 50 e-scooters of 35-km range | 6.4611e-16 | 1.8988e-15 | .3976 |
| 50 e-scooters of 45-km range | 0.0064019 | 1.7141e-20 | 1.5298e-08 |
| 50 e-scooters of 60-km range | 0.1667 | 1.5521e-09 | 0.051345 |
| 100 e-scooters of 35-km range | 0.0027793 | 1.3687e-20 | 2.0081e-05 |
| 100 e-scooters of 45-km range | 0.34828 | 7.7288e-15 | 0.206 |
| 100 e-scooters of 60-km range | 0.55138 | 0.0026561 | 0.26083 |

## Discussion and Conclusion

Recently, the public interest on micro-mobility modes has been proliferating with rapid pace transforming the transportation systems in many cities. Micro-mobility offers affordable and rapid commute opportunity, saving users from intolerable waiting and the squander of time in congested areas. During the COVID-19 pandemic, surveys showed that more people have shifted to micro-mobility ride-sharing modes making it less affected by COVID-19 compared to other public transportation sectors like buses and trains. In this research, we propose a new business model for the micro-mobility ride-sharing systems, where e-scooters and e-bikes are crowd-sourced. This novel model consists of three entities including the suppliers, the customers, and the management party. The proposed model is believed to offer several advantages features for all classes included in this business strategy. We hypothesis that this type of business-based models is getting popular and being welcomed in public appearance and has the potential to change the micro-mobility sharing system market, if it succeeds.

An Agent-based modelling (ABM) framework was built to control the proposed model using five different rules. We tested the model using a dataset that contains nine million e-scooter trips in Austin, TX. We used two criteria to evaluate the aggregated behaviour of the ABM model when changing the central agent rules over four different scenarios. We used two criterions, the Home Index (HI) and the percentage of satisfied demand (PSD). To compare the results, we used mixed effect Gamma regression model to explain the variability in the four scenarios in terms of the combination of the proposed rules. The Gamma mixed effect model was used to test whether the differences are statistically significant. We ran the agent-based simulation six times using three maximum battery ranges (35, 45, and 60 km) and different number of e-scooters at each origin (50, and 100). Our results show the two criterions were met for the four agent levels, meaning that the business-based e-scooter model achieved the two operation criterions, namely, (1) the e-scooter end up at their home location at the end of day (2) the demand is partially satisfied. Both criterions will improve as we increase the available number of e-scooters and the battery ranges.

A promising application of this novel model could be in a crowded city centre, where employees arrive at their offices early in the morning using e-scooters/e-bikes and they stay until 5:00 pm. These e-scooter/e-bike could be offered for rent during the entire period they are at offices under the constrains that at the end of the office hours the e-scooter/e-bike is returned to the same point or nearby point they were at the beginning of the day and have a residual charge enough for the owner to go back home.

## Author Contributions Statement